\documentclass[epj]{svjour}
\usepackage{graphicx}
\usepackage{amsmath}
\usepackage{amssymb}
\usepackage{xfrac}
\usepackage{hyperref}
\usepackage[usenames,dvipsnames]{xcolor}
\graphicspath{{./fig/}}

\newcommand{\sNN}{s_\mathrm{NN}}

\newcommand{\di}{{\rm d}}

\def\wrho{{\widehat{\rho}}}

\newcommand{\tr}{{\rm tr}}

\newcommand{\p}{{\rm p}}

\newcommand{\be}{\begin{equation}}
\newcommand{\ee}{\end{equation}}                                                                               
\newcommand{\bea}{\begin{eqnarray}}
\newcommand{\eea}{\end{eqnarray}}

\begin{document}

\title{Study of $\Lambda$ polarization in relativistic nuclear collisions at $\sqrt{\sNN} = 7.7$--$200$ GeV}

\author{Iu.~Karpenko\inst{1,2} \and F.~Becattini\inst{1,3}}
%\email{yu.karpenko@gmail.com}\email{becattini@fi.infn.it}
\institute{INFN - Sezione di Firenze, Via G. Sansone 1, I-50019 Sesto Fiorentino (Firenze), Italy \and
Bogolyubov Institute for Theoretical Physics,  Ul.~Metrolohichna, 14-b, 03680 Kiev, Ukraine \and
Universit\'a di Firenze, Via G. Sansone 1, I-50019 Sesto Fiorentino (Firenze), Italy}

\abstract{
We present a calculation of the global polarization of $\Lambda$ hyperons in relativistic Au-Au collisions 
at RHIC Beam Energy Scan range $\sqrt{\sNN}=7.7 - 200$~GeV with a 3+1 dimensional 
cascade + viscous hydro model, UrQMD+vHLLE. Within this model, the mean 
polarization of $\Lambda$ in the out-of-plane direction is predicted to decrease 
rapidly with collision energy from a top value of about 2\% at the lowest energy 
examined. We explore the connection between the polarization signal and thermal 
vorticity and estimate the feed-down contribution to $\Lambda$ polarization due to 
the decay of higher mass hyperons.
}

\maketitle

%*************************************************************************
\section{Introduction}
%*************************************************************************

Particles produced in relativistic heavy ion collisions are expected to be polarized 
in peripheral collisions because of angular momentum conservation. At finite impact 
parameter, the Quark Gluon Plasma (QGP) has a finite angular momentum perpendicular 
to the reaction plane and some fraction thereof may be converted into spin of final 
state hadrons. Therefore, measured particles may show a finite mean {\em global} 
polarization along the angular momentum direction. 

Early estimates of this effect \cite{xnwang1} were based on the general idea that polarized quarks
in the QGP stage of the production process would eventually give rise to polarized hadrons, making 
it possible to predict qualitative features of the final hadrons polarization. It was then proposed
\cite{becapicc1,becapicc2} that polarization can be calculated assuming that the spin degrees of
freedom are at local thermodynamical equilibrium at the hadronization stage in much the same way as
the momentum degrees of freedom. In other words, polarization can be predicted finding the 
appropriate extension of the familiar Cooper-Frye formula to particles with spin. A specific derivation 
was presented in refs.~\cite{becapicc1,becaspin} where it was pointed out that the hydrodynamical 
quantity steering the polarization is the {\em thermal vorticity}, that is (minus) the antisymmetric 
part of the gradient of the four-temperature field $\beta = (1/T) u$ where $T$ is the proper temperature 
and $u$ the hydrodynamic velocity:
\be\label{thvort}
   \varpi_{\mu\nu} = -\frac{1}{2} \left( \partial_\mu \beta_\nu - \partial_\nu \beta_\mu \right)
\ee
Particularly, the first-order expansion of the polarization in terms
of thermal vorticity was obtained in ref.~\cite{becaspin} for hadrons with spin \sfrac{1}{2} 
(lately recovered with a different method in ref.~\cite{xnwang2}), yet its extension to higher spins 
could be derived from the corresponding global equilibrium expression \cite{becapicc1}. 

This theoretical work made it possible to make definite quantitative predictions of global $\Lambda$ 
polarization in nuclear collisions from hydrodynamic calculations, with a resulting mean value ranging 
from some permille to some percent~\cite{becacsernai,beca2015,xnwang3}, with an apparently strong 
dependence on the initial conditions, particularly on the initial longitudinal velocity field. Calculations of 
vorticity in relativistic heavy ion collisions - which could be then turned into a polarization map -
were also recently presented in ref.~\cite{liao,deng}. To complete the theoretical overview on the subject, 
it should be pointed out that different approaches, as well as additional mechanisms, to the 
$\Lambda$ polarization in relativistic nuclear collisions were proposed in refs.~\cite{ayala,celso,teryaev,aristova,huang}.

From the experimental viewpoint, while the early measurement by the STAR experiment in Au-Au collisions at 
$\sqrt{s_{\rm NN}}=200$~GeV \cite{Abelev:2007zk} only set an upper limit on the $\Lambda$ polarization
of 0.02, a first strong evidence of non-vanishing global polarization showing the predicted features within the 
hydrodynamical model (orthogonal to the reaction plane, equal sign for particle and antiparticle) has been recently 
reported by the STAR experiment \cite{MLisaTalk,upsal} at the lower energies of the RHIC Beam Energy Scan 
(BES), with a top value at the lowest energy point $\sqrt{s_{\rm NN}} = 7.7$ GeV of some percent.  

This finding certainly opens a new direction in the physics of Quark Gluon Plasma and relativistic heavy
ion collisions and, in the short run, calls for new numerical calculations of thermal vorticity and 
$\Lambda$ polarization in the BES energy range, what is the primary purpose of this work. 
As has been mentioned, previous quantitative estimates of $\Lambda$ polarization in hydrodynamic models 
vary from some permille to some percent, with strong dependence on the collision model and, specifically,
on the hydrodynamic initial conditions. In \cite{becacsernai}, simulation of noncentral 
$\sqrt{s_{\rm NN}}=200$~GeV Au-Au collisions at RHIC with initial state from Yang-Mills dynamics followed 
by a 3~dimensional ideal hydro expansion implying an initial non-vanishing space vorticity vector 
\cite{csernaivort} resulted in polarization of low-$p_T$ $\Lambda$ of few percent magnitude, parallel 
to the direction of total angular momentum of the fireball, whereas for few GeV $p_T$ it reaches about 8\%. 
In ref.~\cite{beca2015}, at the same collision energy, a 3-dimensional hydrodynamic expansion with 
initial state from optical Glauber model with parametrized space-time rapidity dependence chosen to reproduce the momentum rapidity distributions of hadrons and directed 
flow results \cite{Bozek:2010bi} resulted in much lower values of polarization: about 0.2\% for low-$p_T$ 
$\Lambda$ and up to 1.5\% for the longitudinal (along the beam) component of polarization of high-$p_T$ 
$\Lambda$. In a recent work \cite{xnwang3} an event-by-event 3-dimensional viscous hydrodynamics with 
initial state from AMPT model results in similar few permille average global polarization for A+A collisions 
at $\sqrt{\sNN}=62.4, 200$ and $2760$~GeV and the correlation of polarizations of $\Lambda$ pairs is studied.

Herein, we present the simulations of Au-Au collisions at the RHIC BES energies, $\sqrt\sNN=7.7\dots200$~GeV using a 3-dimensional viscous hydrodynamic model vHLLE with initial state from the UrQMD model. Such hybrid approach coupled to final state UrQMD has been tuned to reproduce the basic hadron observables in relativistic heavy ion collisions, that is (pseudo)rapidity and transverse momentum distributions and  elliptic flow coefficients. We calculate the polarization of $\Lambda$ hyperons produced out of the fluid (so-called thermal $\Lambda$). Besides, we estimate, for the first time, the 
contribution to the $\Lambda$ polarization stemming from the decays of likewise polarized $\Sigma^0$, 
$\Sigma(1385)$, as well as other $\Lambda$ and $\Sigma$ resonances up to $\Sigma(1670)$.

The paper is organized as follows: in Section~\ref{sect-model-desc} we briefly describe the model 
used to simulate nuclear collisions in the examined energy range; in Section~\ref{poladef} we summarize 
the main definitions concerning spin and polarization in the relativistic regime, as well as the used 
formulae to calculate the polarization; in Section~\ref{sect-results} we present 
the results for the polarization of $\Lambda$ baryons in Au-Au collisions at BES 
energies and discuss their interpretation in terms of vorticity patterns in the 
hydrodynamic expansion; in Section ~\ref{feeddown} we estimate the corrections due 
to the resonance decays. Conclusions are drawn in Section~\ref{sect-conclusions}.

%*************************************************************************************
\section{Nuclear collision model description}\label{sect-model-desc}
%*************************************************************************************

The full cascade+viscous hydro+cascade model used in our studies is described in detail in ref.~\cite{Karpenko:2015xea}, 
herein we only summarize its main features. At lower collision energies the colliding nuclei do not look 
like thin ``pancakes'' because of weaker Lorentz contraction; also, partonic models of the initial state 
(CGC, IP-Glasma) gradually lose their applicability in this regime. The longitudinal boost invariance is 
not a good approximation anymore, therefore one needs to simulate a 3-dimensional hydrodynamic expansion. 
Such considerations motivate us to choose UrQMD model to describe the dynamics of the initial state from the 
first nucleon-nucleon collisions until a hypersurface of constant Bjorken proper time $\tau=\sqrt{t^2-z^2}=\tau_0$ 
to provide a 3 dimensional initial state for subsequent hydrodynamic stage\footnote{Choice of initial 
hypersurface $\tau=\tau_0$ and $\tau$ as an evolution parameter at the fluid stage is made to keep the 
applicability at higher end of the BES region, where nuclei start to look like thin pancakes.}. For most 
of the collision energies, the value of $\tau_0$ is chosen to correspond to an average time when the two 
incoming nuclei have passed through each other, $\tau_0 = 2R/(\gamma v_z)$.

At the $\tau=\tau_0$ hypersurface the transition to a fluid description occurs: energies and momenta 
of particles crossing the hypersurface are distributed to the hydrodynamic cells around the positions 
of particles according to a Gaussian profile. The contribution of each particle to a fluid cell
$\{i,j,k\}$ is given by:
\begin{align}
\left\{ \Delta P^\mu_{ijk}, \Delta N^0_{ijk} \right\}=\left\{ P^\mu, N^0 \right\} \cdot C\cdot\qquad & \nonumber
\\\cdot\exp\left(-\frac{\Delta x_i^2+\Delta y_j^2}{R_\perp^2}-\frac{\Delta\eta_k^2}{R_\eta^2}\gamma_\eta^2 \tau_0^2\right).&
\end{align}
where $\Delta x_i$, $\Delta y_j$, $\Delta \eta_k$ are the differences between particle's position 
and coordinates of the center of the hydrodynamic cell $\{i,j,k\}$ in the transverse plane and space-time rapidity, $R_\perp$ and $R_\eta$ are the coarse-graining (smearing) parameters and $\gamma_\eta={\rm cosh}(y_p-\eta)$ is the
longitudinal Lorentz factor of the particle as seen in a frame moving with the rapidity $\eta$; the normalization constant $C$ is calculated in order to conserve energy/momentum in this transition.

With such initial conditions the 3-dimensional viscous hydrodynamic evolution in the Israel-Stewart formulation starts, which is numerically solved with \texttt{vHLLE} code \cite{Karpenko:2013wva}. In particular, we work in Landau frame, and neglect baryon and electric charge diffusion currents. Whereas the bulk viscosity is set to zero, $\zeta/s=0$, for the shear-stress tensor $\pi^{\mu\nu}$ the following evolution equation is solved:
\begin{equation}
\left<u^\gamma \partial_{;\gamma} \pi^{\mu\nu}\right>
 =-\frac{\pi^{\mu\nu}-\pi_\text{NS}^{\mu\nu}}{\tau_\pi}
  -\frac 4 3 \pi^{\mu\nu}\partial_{;\gamma}u^\gamma, \label{evolutionShear}
\end{equation}
where the relaxation time of the shear-stress tensor is set to $\tau_\pi = 5\eta/(Ts)$, the brackets denote the traceless and orthogonal to $u^\mu$ part of the tensor and $\pi_\text{NS}^{\mu\nu}$ is the Navier-Stokes value
of the shear-stress tensor.

Finally, we set the transition from fluid to particle description 
(a.k.a.\ {\em particlization}) to happen at a certain energy density $\epsilon=\epsilon_{\rm sw}$. 
The elements of particlization hypersurface $\Sigma$ are computed in the course of hydrodynamic 
evolution by means of the \texttt{CORNELIUS} subroutine \cite{Huovinen:2012is}. To calculate 
basic hadronic observables, the Monte Carlo hadron sampling at 
the particlization surface is performed using basic Cooper-Frye formula with standard quadratic ansatz 
for the shear viscous corrections. The sampled hadrons are fed into UrQMD cascade to calculate 
the final state hadronic interactions.

It has been shown in ref.~\cite{Karpenko:2015xea} that in such a model it is not possible to 
reproduce the basic bulk hadronic observables - (pseudo)rapidity distributions, transverse 
momentum spectra and elliptic flow coefficients - simultaneously in the collision energy 
range $\sqrt{s_{\rm NN}}=7.7\dots200$~GeV if the parameters of the model (except the hydro 
starting time $\tau_0$) are fixed to constant values. However, a reasonable reproduction 
of the experimental data has been achieved when the parameters of the model were chosen 
to depend monotonically on the collision energy as it is shown in Table~\ref{tb-params}. 
This was obtained when the particlization energy density was fixed to 
$\epsilon_{\rm sw}=0.5$~GeV/fm$^3$ for the whole collision energy range.
%-----------------------------------------------------------------------------------------
\begin{table}
\vspace{10pt}
\begin{tabular}{|l|l|l|l|l|}
\hline
 $\sqrt{\sNN}$~[GeV] & $\tau_0$~[fm/c] & $R_\perp$~[fm] & $R_\eta$~[fm] & $\eta/s$ \\ \hline
     7.7          &      3.2        &     1.4        &     0.5    &    0.2   \\ \hline
     8.8 (SPS)    &      2.83       &     1.4        &     0.5    &    0.2   \\ \hline
     11.5         &      2.1        &     1.4        &     0.5    &    0.2   \\ \hline
     17.3 (SPS)   &      1.42       &     1.4        &     0.5    &    0.15  \\ \hline
     19.6         &      1.22       &     1.4        &     0.5    &    0.15  \\ \hline
     27           &      1.0        &     1.2        &     0.5    &    0.12  \\ \hline
     39           &      0.9*        &     1.0        &     0.7    &    0.08  \\ \hline
     62.4         &      0.7*        &     1.0        &     0.7    &    0.08  \\ \hline
     200          &      0.4*        &     1.0        &     1.0    &    0.08  \\ \hline
 \end{tabular}
\caption{Collision energy dependence of the model parameters chosen to
  reproduce the experimental data in the RHIC BES range: $\sqrt{\sNN}=7.7-200$~GeV. An asterisk denotes the values of starting time $\tau_0$ which are adjusted instead of being taken equal to $2R/(\gamma v_z)$.}\label{tb-params}
\end{table}
%----------------------------------------------------------------------------------------------

The UrQMD cascade does not treat polarization of hadrons, therefore we calculate the polarization of particles produced at the particlization surface only. To to so, have replaced the Monte Carlo hadron sampling with a direct calculation based on the eq.~(\ref{eq-Pip}) applied on particlization surfaces from event by event hydrodynamics. The final state hadronic cascade (UrQMD) is not used in the present study.

%*************************************************************************************
\section{Spin and polarization}\label{poladef}
%*************************************************************************************

We summarize in this section the basic definitions concerning massive relativistic particles with spin.

The starting point is the so-called Pauli-Lubanski vector operator $\widehat{W}^\mu$ which is
defined as follows:
\begin{equation}
 \widehat{W}^\mu = -\frac{1}{2} \epsilon^{\mu\nu\rho\sigma} \widehat{J}_{\nu\rho} \widehat{P}_\sigma
\end{equation}
where $\widehat{J}$ and $\widehat{P}$ are the angular momentum-boost operators and four-momentum
operators for a single particle. The proper spin four-vector operator is simply the adimensional 
vector obtained by dividing $\widehat{W}$ by the mass, that is:
\begin{equation}
 \widehat{S}^\mu = \frac{1}{m} \widehat{W}^\mu
\end{equation}
It can be easily shown \cite{wukitung} that:
$$
  [\widehat{S}^\mu,\widehat{P}^\nu] = 0  \qquad \qquad \widehat{S}^\mu \widehat{P}_\mu = 0
$$
According to the latter relation, the spin vector is spacelike and has only three independent
components. Therefore, for single particle states with definite four-momentum $p$ it can be decomposed 
\cite{moussa} along three spacelike vectors $n_i(p)$ with $i=1,2,3$ orthogonal to $p$:
\begin{equation}\label{spindec}
  \widehat{S}^\mu = \sum_{i=1}^3 \widehat{S}_i(p) n_i(p)^\mu
\end{equation}
It can be shown that the operators $\widehat{S}_i(p)$ with $i=1,2,3$ obey the well 
known SU(2) commutation relations and they are indeed the generators of the little 
group, the group of transformations leaving $p$ invariant for a massive particle. In 
other, maybe simpler, words they correspond to the familiar spin operators of the 
non-relativistic quantum mechanics in the particle's rest frame. Also, the 
$\widehat{S}^\mu\widehat{S}_\mu$ operator is a Casimir of the full Poincar\'e group 
whose eigenvalue is $S(S+1)$ where $S$ is {\em the} spin of the particle.

We denote the {\em mean spin vector} the four-vector obtained by taking both the quantum 
and thermal average of $\widehat{S}$, that is:
\begin{equation}
 S^\mu = \langle \widehat{S}^\mu \rangle \equiv \tr (\wrho \widehat{S}^\mu )
\end{equation}
and each component takes on values in the interval $(-S,S)$. Indeed, the properly 
called polarization vector is obtained by normalizing $\Pi^\mu$ to the spin of the 
particle, that is:
\begin{equation}\label{polaspin}
 P^\mu = \frac{1}{S} S^\mu
\end{equation}
hence each component takes on values in the interval $(-1,1)$. 

For a multi-particle system, the calculation of the polarization of particles with 
four-momentum $p$ requires the decomposition of the mean total angular momentum into 
momentum modes. In general, this requires the knowledge of the Wigner function, which 
allows to express the mean values of operators as integrals over space-time and four-momentum 
space. By using such theoretical tools, the mean spin vector of spin $1/2$ particles with 
four-momentum $p$, produced around point $x$ on particlization hypersurface, at the
leading order in the thermal vorticity reads~\cite{becaspin}:
\begin{equation}\label{eq-Pixp}
  S^\mu(x,p)= - \frac{1}{8m} (1-f(x,p)) \epsilon^{\mu\nu\rho\sigma} p_\sigma \varpi_{\nu\rho}
\end{equation}
a formula recovered in ref.~\cite{xnwang2}. In eq.~(\ref{eq-Pixp}) $f(x,p)$ is the Fermi-Dirac distribution 
and $\varpi$ is given by eq.~\ref{thvort} at the point $x$.

In hydrodynamic picture of heavy ion collisions, particles with a given momentum are 
produced across entire particlization hypersurface. Therefore to calculate the relativistic 
mean spin vector of a given particle species with given momentum, one has to integrate 
the above expression over the particlization hypersurface $\Sigma$ \cite{becaspin}:
\begin{equation}\label{eq-Pip}
 S^\mu(p)=\frac{\int d\Sigma_\lambda p^\lambda f(x,p) S^\mu(x,p)}{\int d\Sigma_\lambda 
 p^\lambda f(x,p)}
\end{equation}
The mean (i.e.\ momentum average) spin vector of all particles of given species  can be expressed as:
\begin{equation}\label{eq-Pi}
  S^\mu =\frac{1}{N} \int\frac{\di^3 \p}{p^0} \int d\Sigma_\lambda p^\lambda f(x,p) S^\mu(x,p)
\end{equation}
where $N=\int\frac{\di^3\p}{p^0}\int d\Sigma_\lambda p^\lambda f(x,p)$ is the average number of 
particles produced at the particlization surface.

In the experiment, the $\Lambda$ polarization is measured in its rest frame, therefore one can derive 
the expression for the mean polarization vector in the rest frame from (\ref{eq-Pi}) taking into account 
Lorentz invariance of most of the terms in it:
\begin{align}
  S^{*\mu} = \frac{1}{N} \int \frac{\di^3 \p}{p^0} \int d\Sigma_\lambda p^\lambda f(x,p) 
   S^{*\mu}(x,p)
\end{align}
where asterisk denotes a quantity in the rest frame of particle.

As has been mentioned, the formula (\ref{eq-Pixp}) applies to spin $1/2$ particles. However, a very
plausible extension to higher spins can be obtained by taking into account that the expression of the
spin vector for massive particle of any spin in the Boltzmann statistics is known at global equilibrium
with rotation \cite{becapicc1}; this is discussed in detail in ref.~\cite{bkluv}. The extension
of the formula (\ref{eq-Pixp}), in the limit of Boltzmann statistics, reads:
\begin{equation}\label{pixpgen}
  S^\mu(x,p) \simeq - \frac{1}{2m} \frac{S(S+1)}{3} \epsilon^{\mu\nu\rho\sigma} p_\sigma \varpi_{\nu\rho}
\end{equation}
%

%*****************************************************************************************
\section{Results}\label{sect-results}
%*****************************************************************************************
%======================================================================================
\subsection{Coordinate system}
%======================================================================================

%----------------------------------------------------------------------------------------------
\begin{figure}
\includegraphics[width=0.5\textwidth]{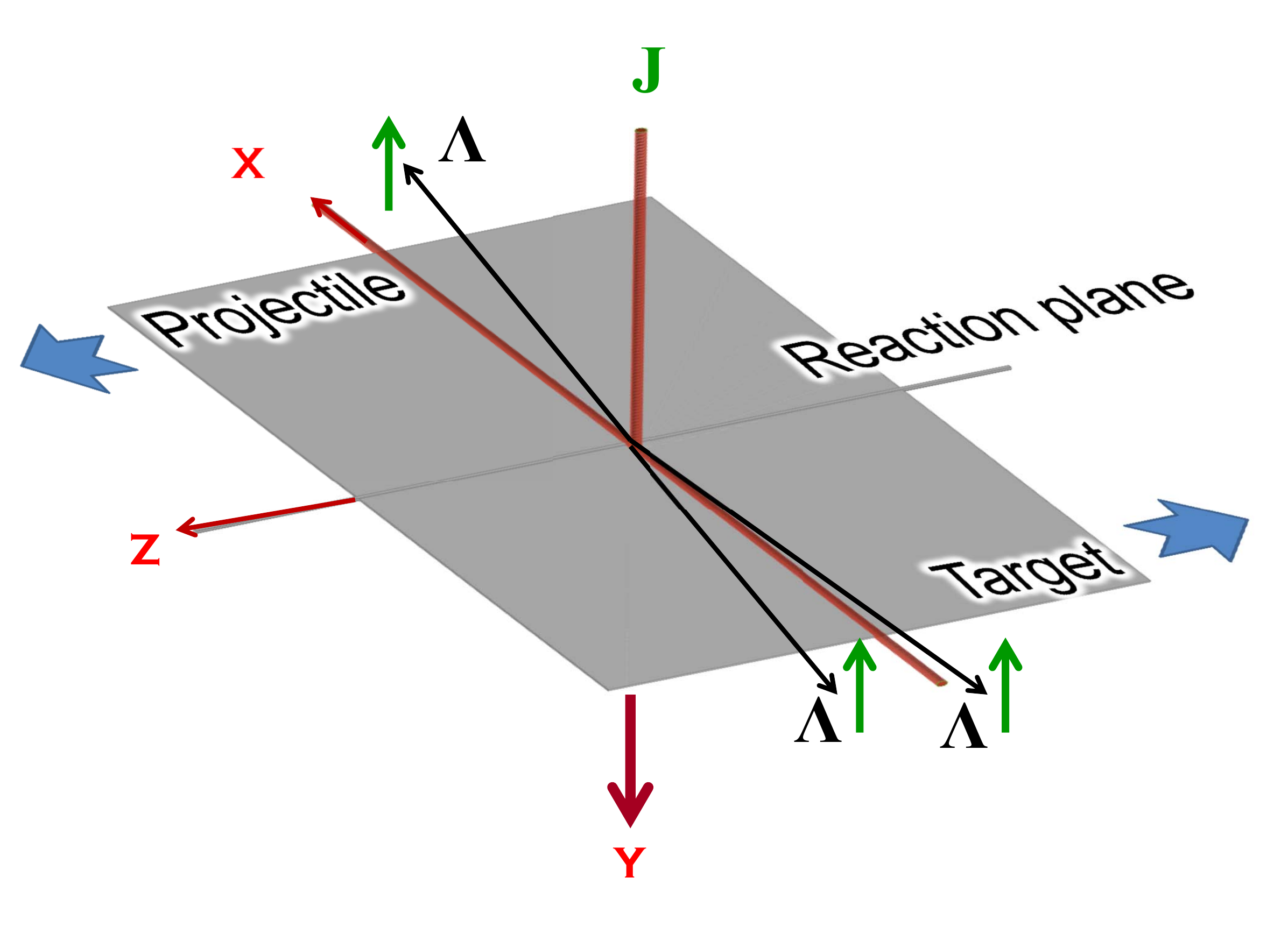}
\caption{Coordinate system used in the calculations. The figure is taken from \cite{becacsernai}.}\label{fig-coord}
\end{figure}
%-----------------------------------------------------------------------------------------------

We use the following notation for the the reference frame in the simulations: $z$ axis points in the 
direction of the beam, and $x$ axis is parallel to the impact parameter. Thus $xz$ is the reaction plane, 
and $y$ axis points out of the reaction plane, oppositely oriented to the total angular momentum ${\bf J}$. 
For the components of the polarization vector (\ref{polaspin}) we use physically motivated aliases:
$P_{||}$ is component in the direction of beam ($z$), $P_b$ is component in direction of impact parameter
($x$) and $P_J$ is component in direction of the total angular momentum of the fireball $J$ (i.e. -$y$).

%======================================================================================
\subsection{Polarization of thermal $\Lambda$ as a function of transverse momentum}
%======================================================================================

\begin{figure*}
\includegraphics[width=\textwidth]{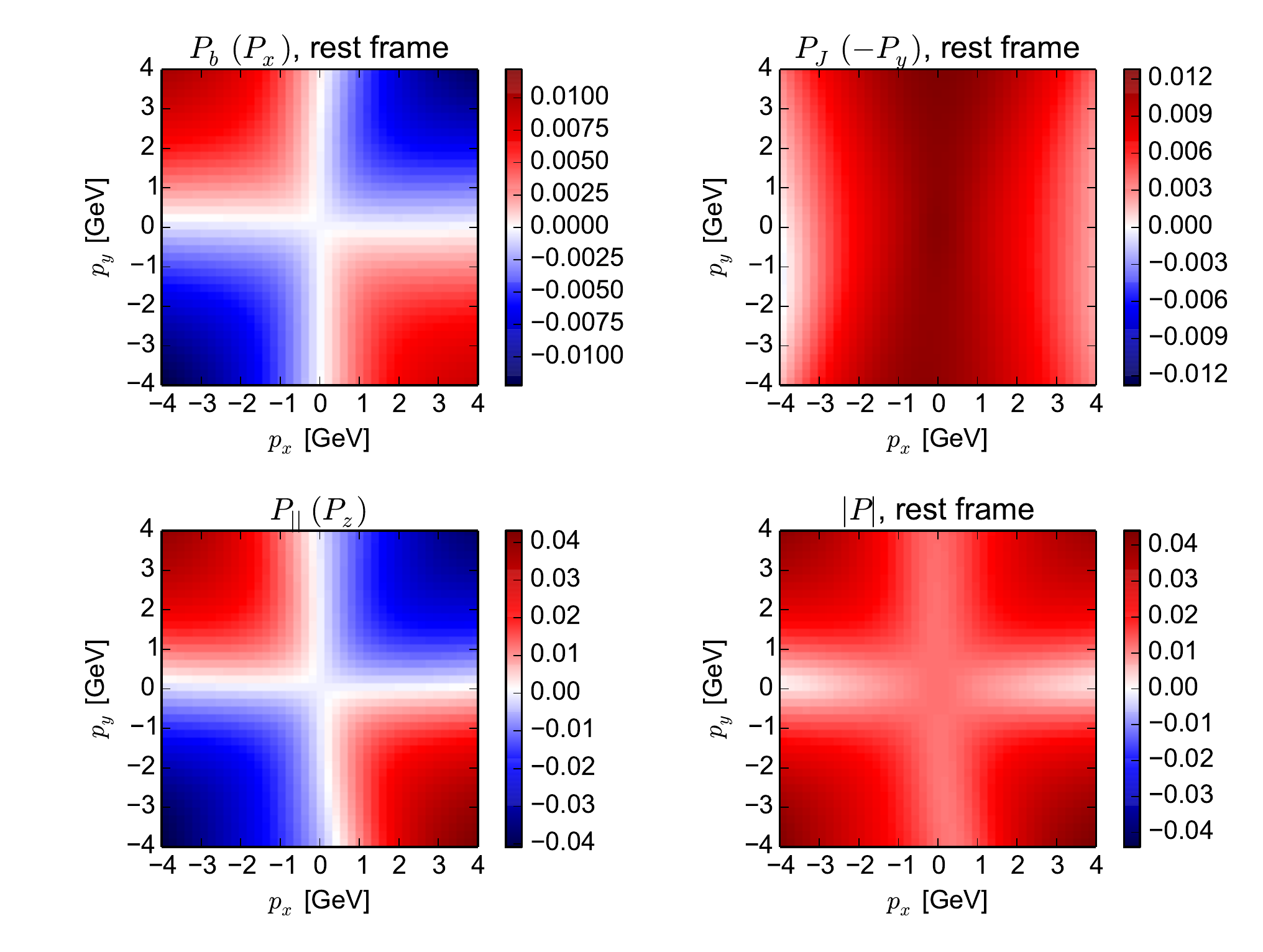}
\caption{Components of mean polarization vector of primary $\Lambda$ baryons produced at zero momentum space rapidity, 
calculated in the model for 40-50\% central Au-Au collisions at $\sqrt{s_{\rm NN}}=19.6$~GeV. 
The polarization is calculated in the rest frame of $\Lambda$.}\label{fig-196gev-4050}
\end{figure*}

We started simulating semi-central Au-Au collisions at one particular collision energy in the middle 
of the Beam Energy Scan range, $\sqrt{s_{\rm NN}}=19.6$~GeV. We ran 1000 event-by-event hydro calculations 
with initial state corresponding to 40-50\% centrality class (impact parameter range $b=9.3-10.4$~fm) 
and calculated event-averaged denominator and numerator of eq.~\ref{eq-Pip} in the $p_x p_y$ plane at zero momentum space rapidity. The resulting components of momentum differential polarization vector (see eq.~(\ref{polaspin}) for definition)
are shown on Fig.~\ref{fig-196gev-4050}. 
One can see that the ${\bf p}_T$ dependence of the polarization has a pattern which is similar to the 
one obtained in a 3+1D hydrodynamic simulation of $\sqrt{s_{\rm NN}}=200$~GeV Au-Au system with 
\texttt{ECHO-QGP} code \cite{echoqgp,beca2015}, including the signs of polarization in different quadrants of 
the $p_x p_y$ plane. On this plot, we extend the calculation up to $p_\perp\approx$5.5~GeV, which is well beyond the applicability of hydrodynamics, to show the trends in the momentum dependence. $\Lambda$ production decreases exponentially with $p_\perp$, therefore larger polarization of high-$p_\perp$ $\Lambda$ does not influence the momentum-averaged polarization.
%------------------------------------------------------------------------------------------------------
%-------------------------------------------------------------------------------------------------------

\begin{figure*}
\includegraphics[width=0.49\textwidth]{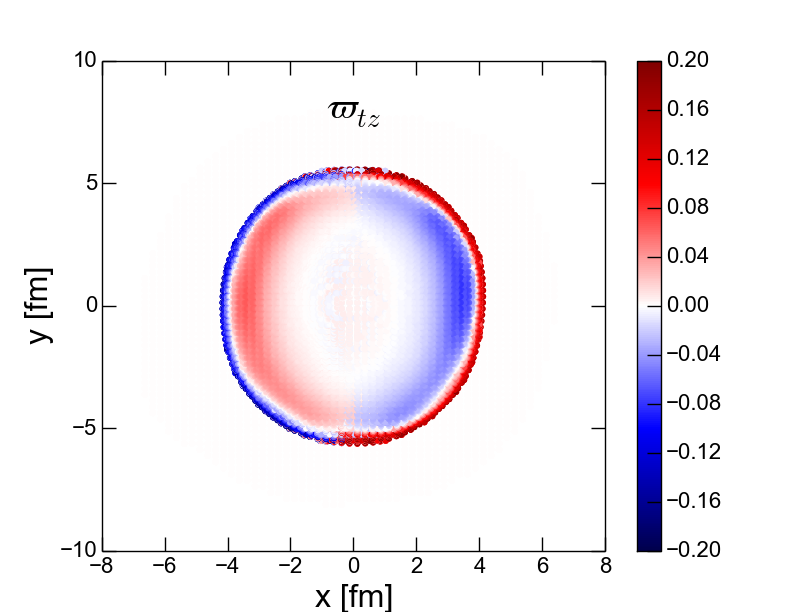}
\includegraphics[width=0.49\textwidth]{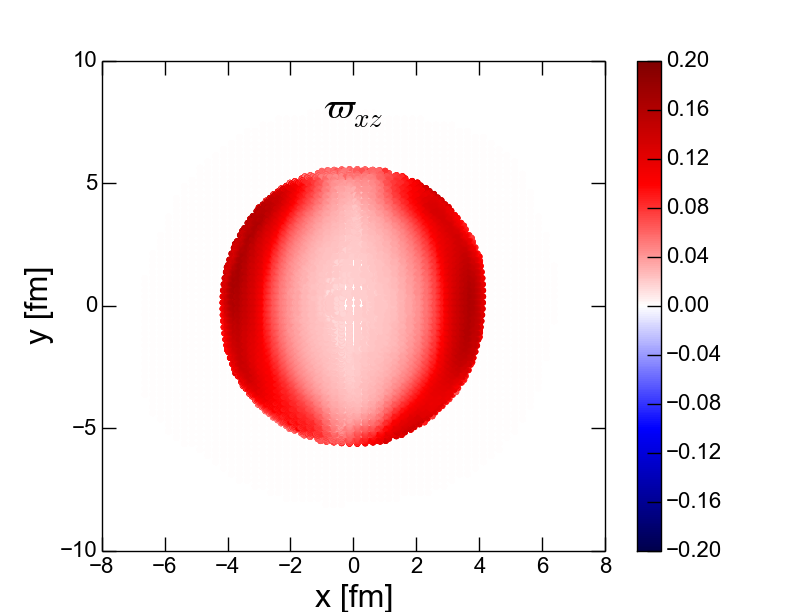}
\caption{Components of thermal vorticity $\varpi_{tz}$ (left) and $\varpi_{xz}$ (right) on the zero space-time
rapidity slice of particlization hypersurface, projected on the $xy$ plane.}\label{fig-omega-196gev}
\end{figure*}

The polarization patterns in $p_x p_y$ plane reflect the corresponding patterns of the components 
of thermal vorticity over the particlization hypersurface. In particular, we found that the leading 
contribution to $P_b$ stems from the term $\varpi_{tz}p_y$ in Eq.~\ref{eq-Pixp}. In turn, 
$\varpi_{tz}$, shown in left panel of Fig.~\ref{fig-omega-196gev} is a result of the interplay 
of $\partial_t \beta_z$ (acceleration of longitudinal flow and temporal gradients of temperature 
- conduction) and $\partial_z \beta_t$ (convection and conduction), according to eq.~(\ref{thvort}).
The $P_J$ component has a leading contribution from the term $\varpi_{xz}p_0$ (which is also 
the only non-vanishing contribution at $p_T=0$), and $\varpi_{xz}$ has a rather uniform profile 
over the zero space-time rapidity slice of the particlization hypersurface, and the leading contribution to it 
comes from $\partial_x u_z$ (shear flow in $z$ direction).
%------------------------------------------------------------------------------------
%--------------------------------------------------------------------------------------

At large transverse momenta, the $P_{||}$ component of polarization has the largest value; however,
since $P^b$ and $P_{||}$ flip signs in different quadrants in $p_x p_y$ plane, their $p_T$ integrated values vanish,
and the only nonzero component remaining is $P_J$, which is parallel to the direction of 
total angular momentum $\bf J$ of the fireball. This goes in line with physical expectation that the global polarization has to be collinear with the vector of the total angular momentum of the system.

%======================================================================================
\subsection{Collision energy dependence of the polarization}
%======================================================================================
%-------------------------------------------------------------------------------------------
\begin{figure}
\includegraphics[width=0.49\textwidth]{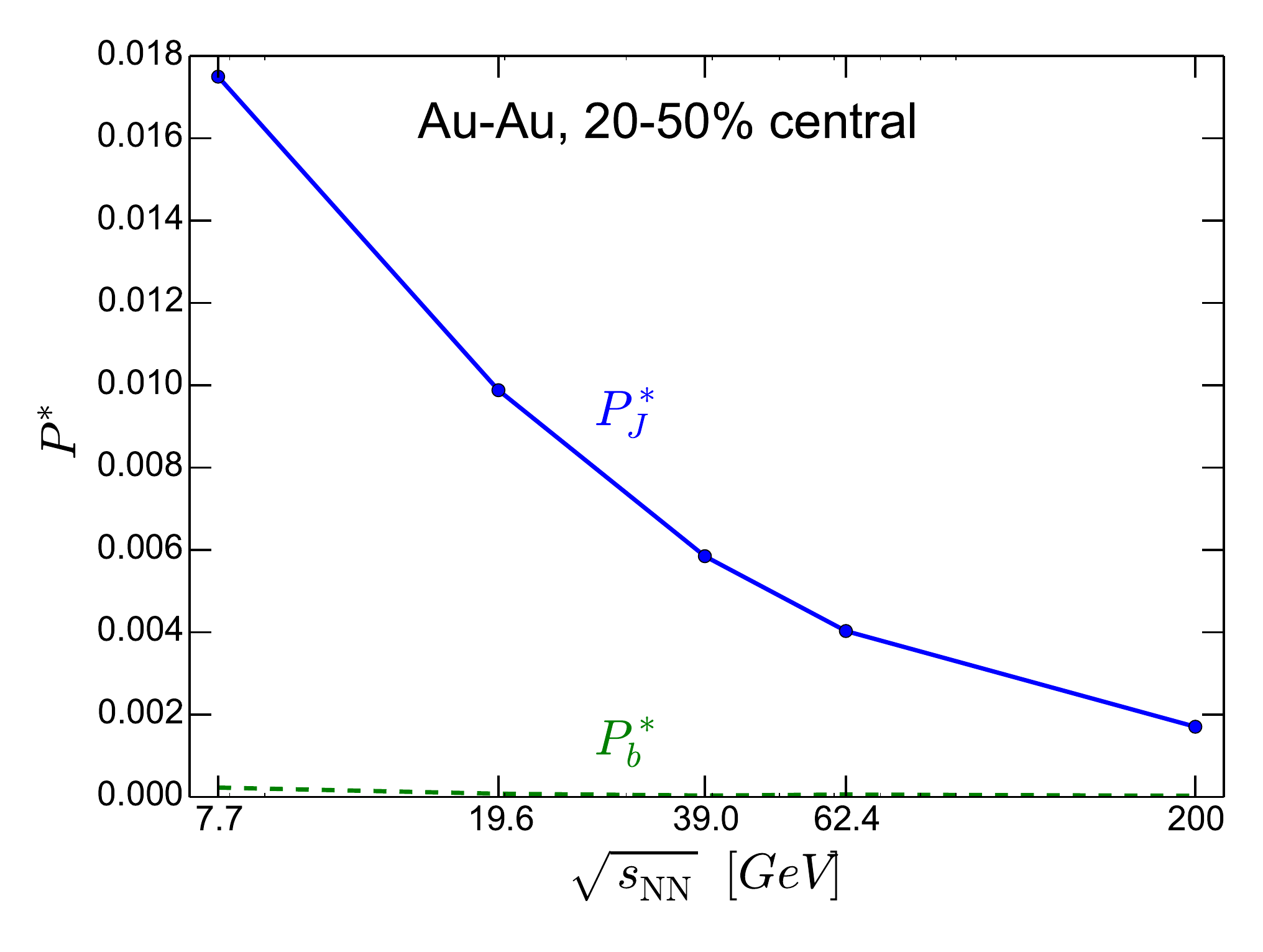}
\caption{Collision energy dependence of the components of polarization vector of $\Lambda$, 
calculated in its rest frame, calculated in the model for 20-50\% central Au-Au collisions.}\label{fig-Pixy-sqrts}
\end{figure}
%-------------------------------------------------------------------------------------------
\begin{figure}
\includegraphics[width=0.5\textwidth]{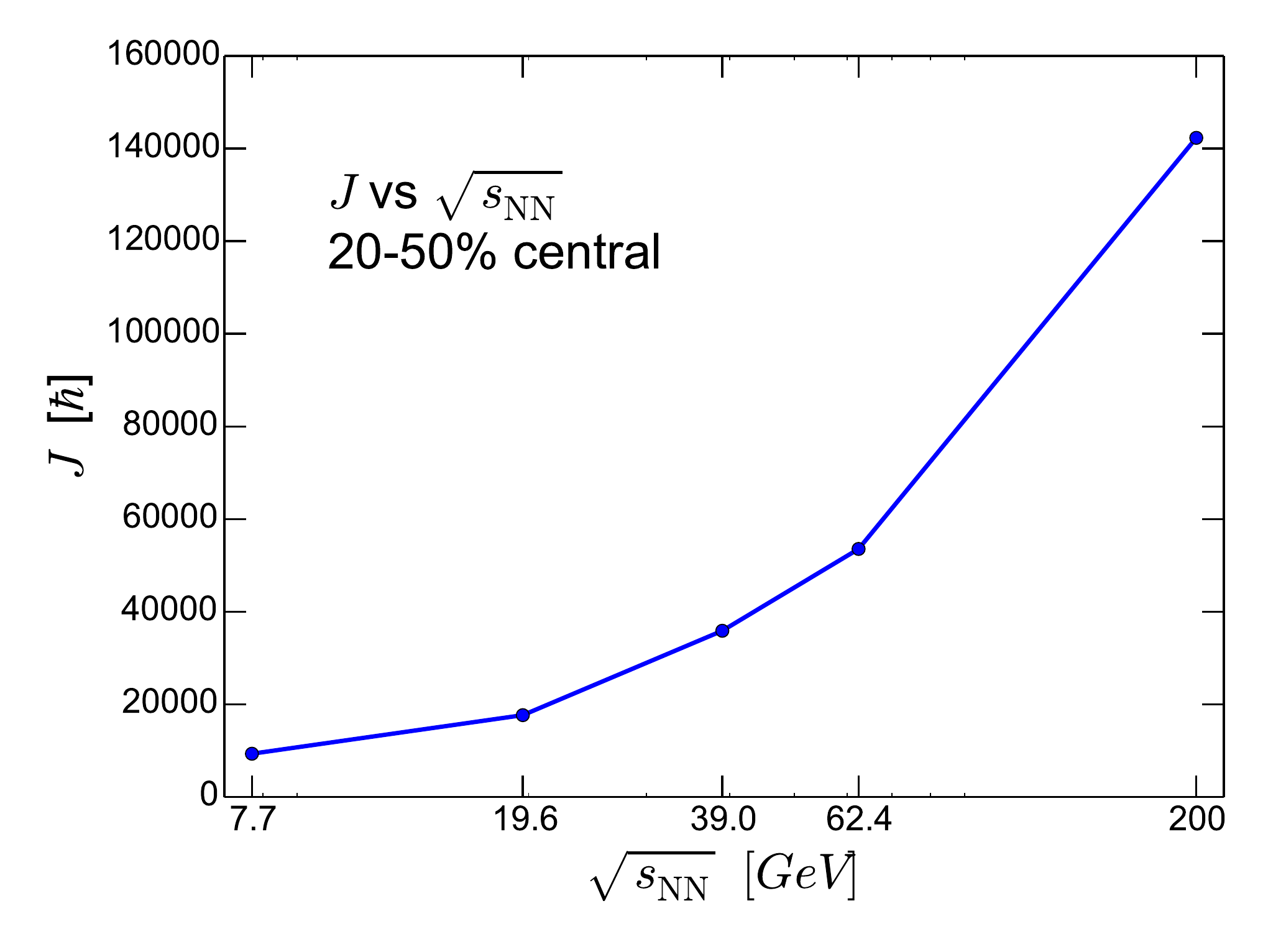}
\includegraphics[width=0.5\textwidth]{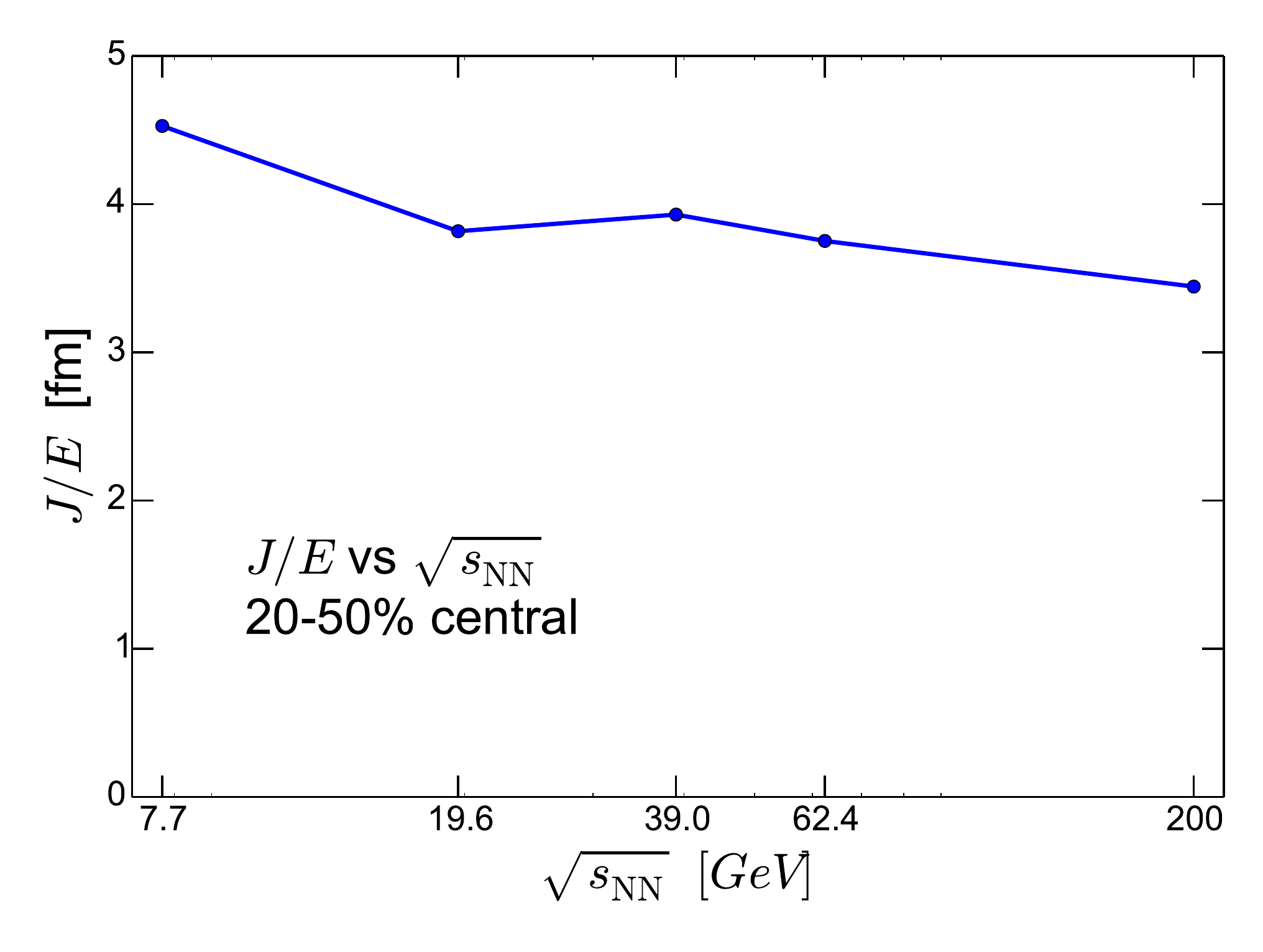}
\caption{Total angular momentum of the fireball (left) and total angular momentum scaled by 
total energy of the fireball (right) as a function of collision energy, calculated in the model 
for 20-50\% central Au-Au collisions.}\label{fig-Jy-sqrts}
\end{figure}
%-------------------------------------------------------------------------------------------
Next we ran the simulations for the full BES collision energy range. To follow recent STAR 
measurements, we choose 20-50\% centrality bin by correspondingly chosen range of impact 
parameters for the initial state UrQMD calculation. The resulting ${\bf p}_T$ integrated 
polarization is shown in Fig.~\ref{fig-Pixy-sqrts}. We observe that the mean polarization 
component along $J$, that is $P_J$ decreases by about one order of magnitude as collision 
energy increases from $\sqrt{s_{\rm NN}}=7.7$~GeV to full RHIC energy, where it turns out 
to be consistent with the results of \cite{beca2015}. 

The fall of the out-of-plane component $P_J$ is not directly related to a change in the 
out-of-plane component of total angular momentum of the fireball. In fact, the total 
angular momentum increases as the collision energy increases, which can be seen on top 
panel of Fig.~\ref{fig-Jy-sqrts}. However, the total angular momentum is not an intensive 
quantity like polarization, so, to have a better benchmark we took the ratio between 
the total angular momentum and the total energy, $J/E$, which is shown in the bottom 
panel of the same figure. Yet, one can see that the $J/E$ shows only a mild decrease 
as collision energy increases.
%-------------------------------------------------------------------------------------------
\begin{figure}
\includegraphics[width=0.5\textwidth]{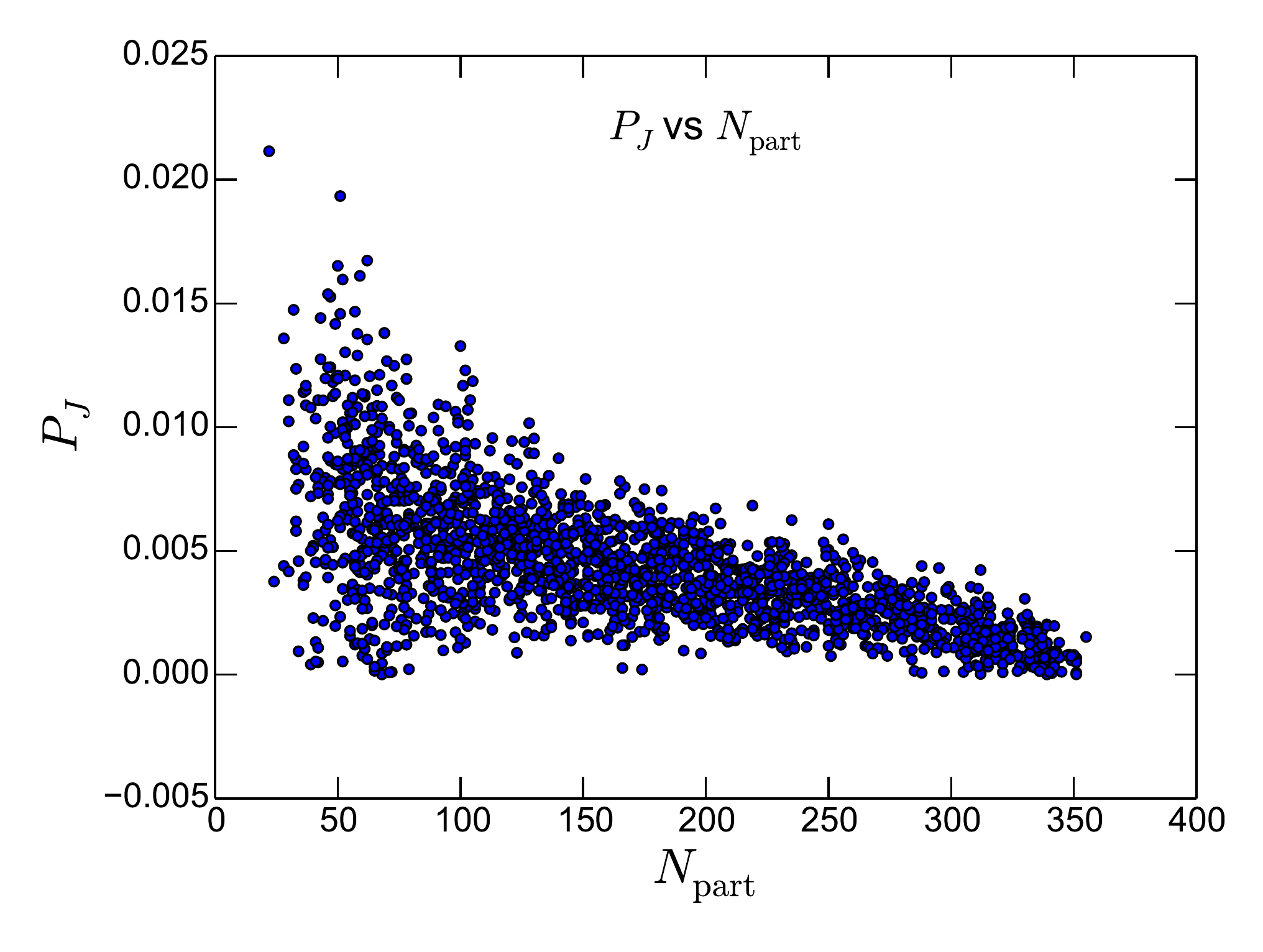}
\includegraphics[width=0.5\textwidth]{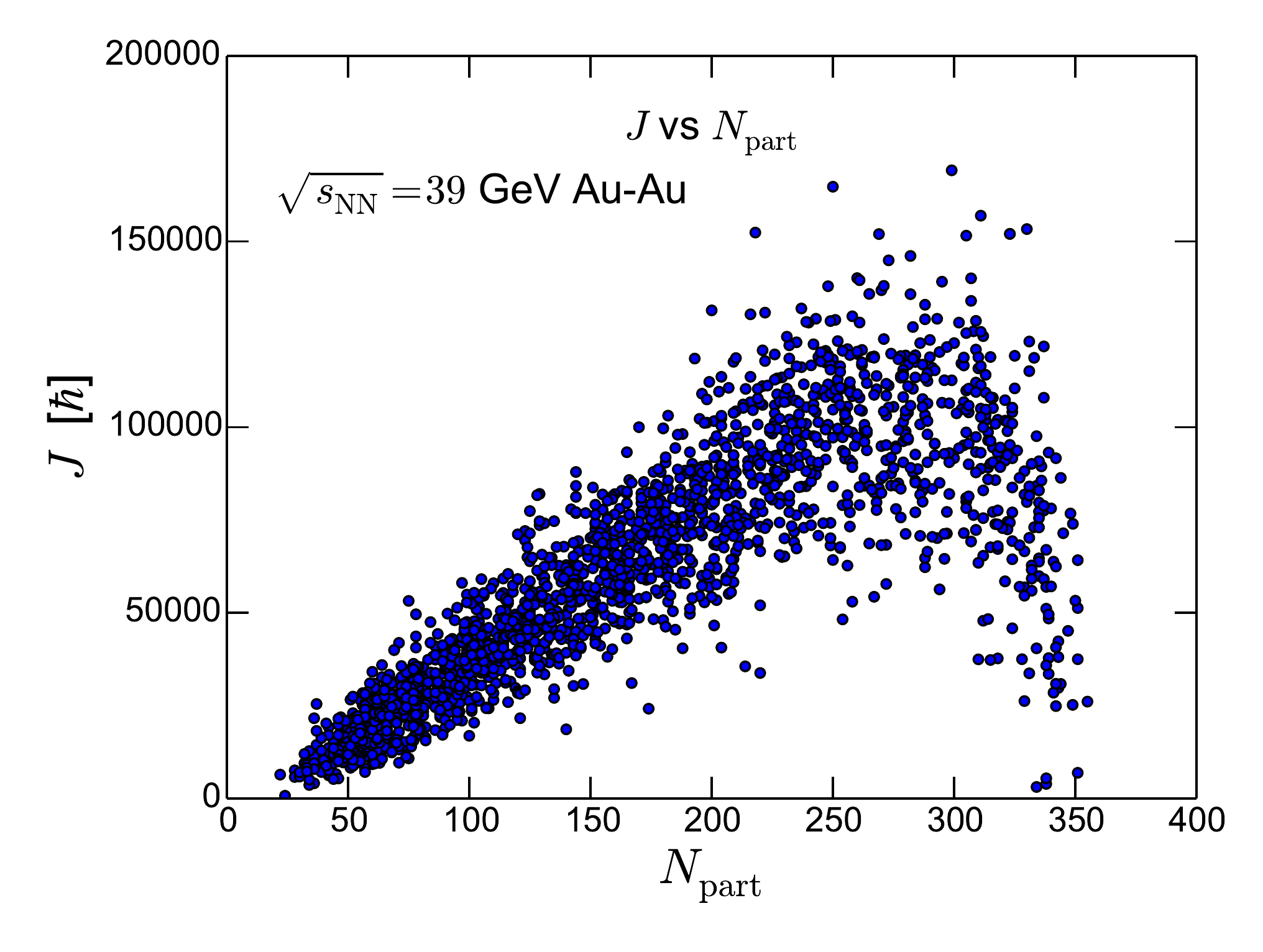}
\caption{Top: out-of-plane $\Lambda$ polarization component as a function of the number 
of participating nucleons $N_{\rm part}$ in particular initial state. Each point represents 
one hydrodynamic configuration in the ensemble of 2000 event-by-event calculations for 
0-50\% central Au-Au collisions at $\sqrt{s_{\rm NN}}=39$~GeV. Bottom: out-of-plane 
component of initial angular momentum versus number of participating nucleons $N_{\rm part}$ 
in the same calculation.}\label{fig-Jy-npart}
\end{figure}
%-------------------------------------------------------------------------------------------

In Fig.~\ref{fig-Jy-npart} we show the distribution of the average polarization of $\Lambda$ 
as a function of centrality (i.e. $N_{\rm part}$), where each point corresponds to a 
hydrodynamic evolution with a given fluctuating initial condition characterized by 
$N_{\rm part}$; in the bottom panel one can see the corresponding distribution of total angular 
momentum $J$. We observe that the total angular momentum distribution has a maximum at 
certain range of $N_{\rm part}$, and drops to zero for the most central events (where 
the impact parameter is zero) and most peripheral ones (where the system becomes small). 
In contrast to that, the polarization shows a steadily increasing trend towards 
peripheral collisions, where it starts to fluctuate largely from event to event because 
of smallness of the fireball, a situation where the initial state fluctuations start to 
dominate in the hydrodynamic stage.
%-------------------------------------------------------------------------------------------
\begin{figure}
\includegraphics[width=0.5\textwidth]{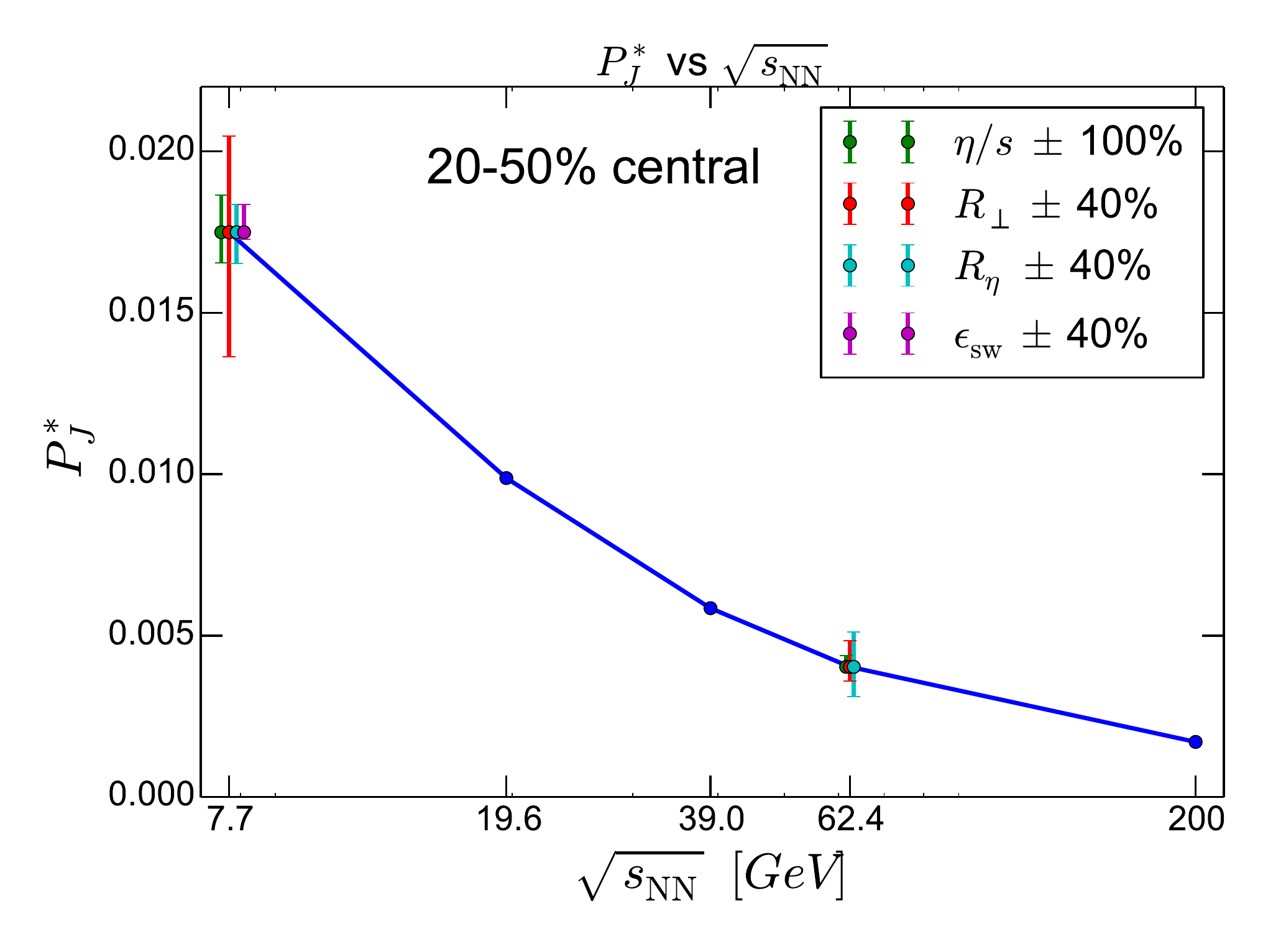}
\caption{Same as Fig.~\ref{fig-Pixy-sqrts} but with bands added, which correspond to 
variations of the model parameters.}\label{fig-Piy-errorbars}
\end{figure}
%-------------------------------------------------------------------------------------------

As it has been mentioned above, the parameters of the model are taken to monotonically 
depend on collision energy in order to approach the experimental data for basic hadronic 
observables. The question may arise whether the collision energy dependence of $P_J$ 
is the result of an interplay of collision energy dependencies of the parameters. We 
argue that it is not the case: in Fig.~\ref{fig-Piy-errorbars} one can see how the $p_T$ 
integrated polarization component $P_J$ varies at two selected collision energies, 
$\sqrt{s_{\rm NN}}=7.7$ and $62.4$~GeV, when the granularity of the initial state 
controlled by $R_\perp$, $R_\eta$ parameters, shear viscosity to entropy ratio of the 
fluid medium $\eta/s$ and particlization energy density $\epsilon_{\rm sw}$ change. 
It turns out that a variation of $R_\perp$ within $\pm 40\%$ changes $P_J$ by $\pm 20\%$, 
and a variation of $R_\eta$ by $\pm 40\%$ changes $P_J$ by $\pm 25\%$ at $\sqrt{s_{\rm NN}}=62.4$~GeV 
only. The variations of the remaining parameters affect $P_J$ much less. We thus conclude 
that the observed trend in $p_T$ integrated polarization is robust with respect to variations
of parameters of the model. 

%======================================================================================
\subsection{Discussion on the energy dependence}
%======================================================================================

%-------------------------------------------------------------------------------------------
\begin{figure*}
\includegraphics[width=0.46\textwidth]{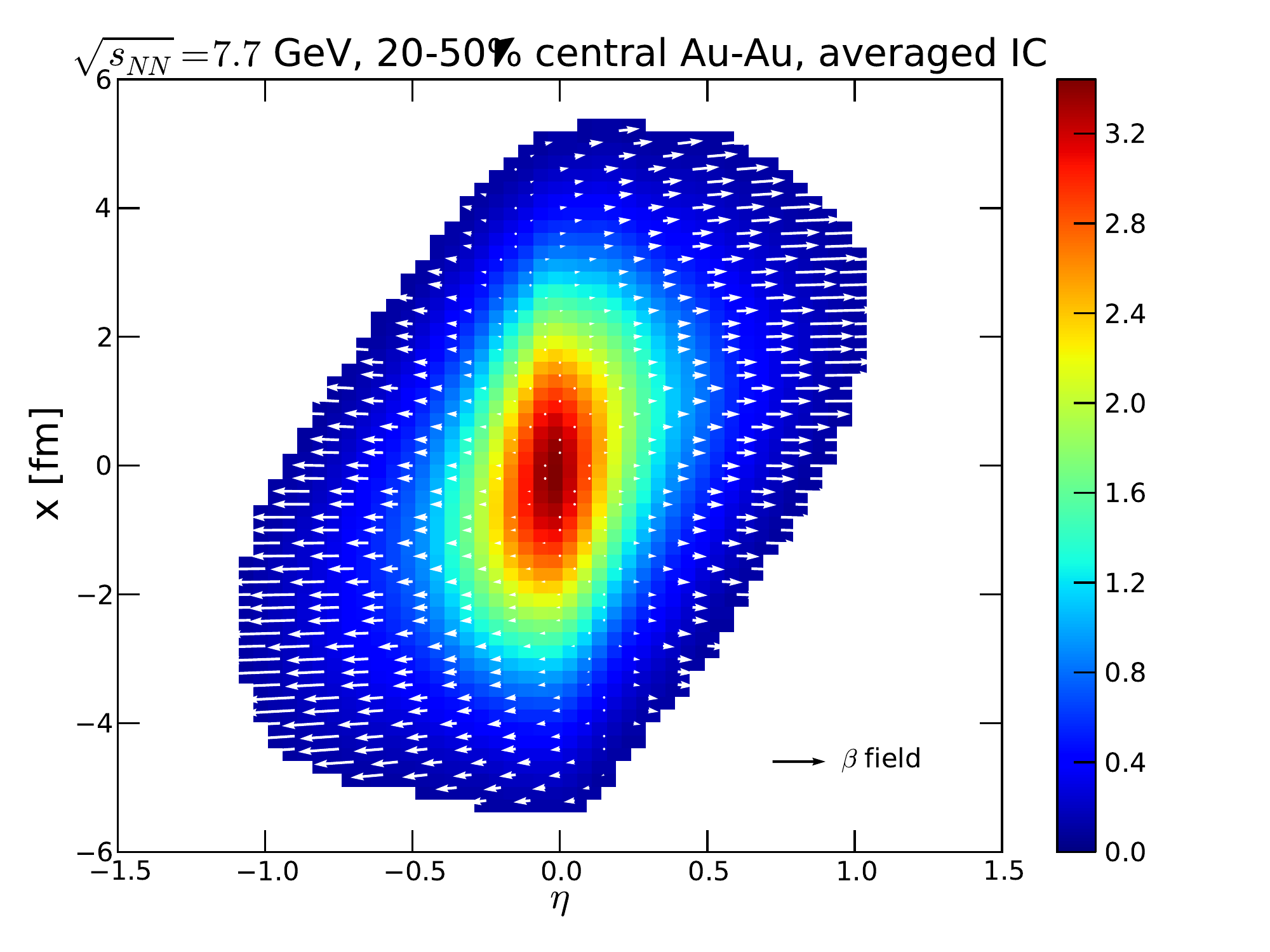}\hspace{-10pt}
\includegraphics[width=0.47\textwidth]{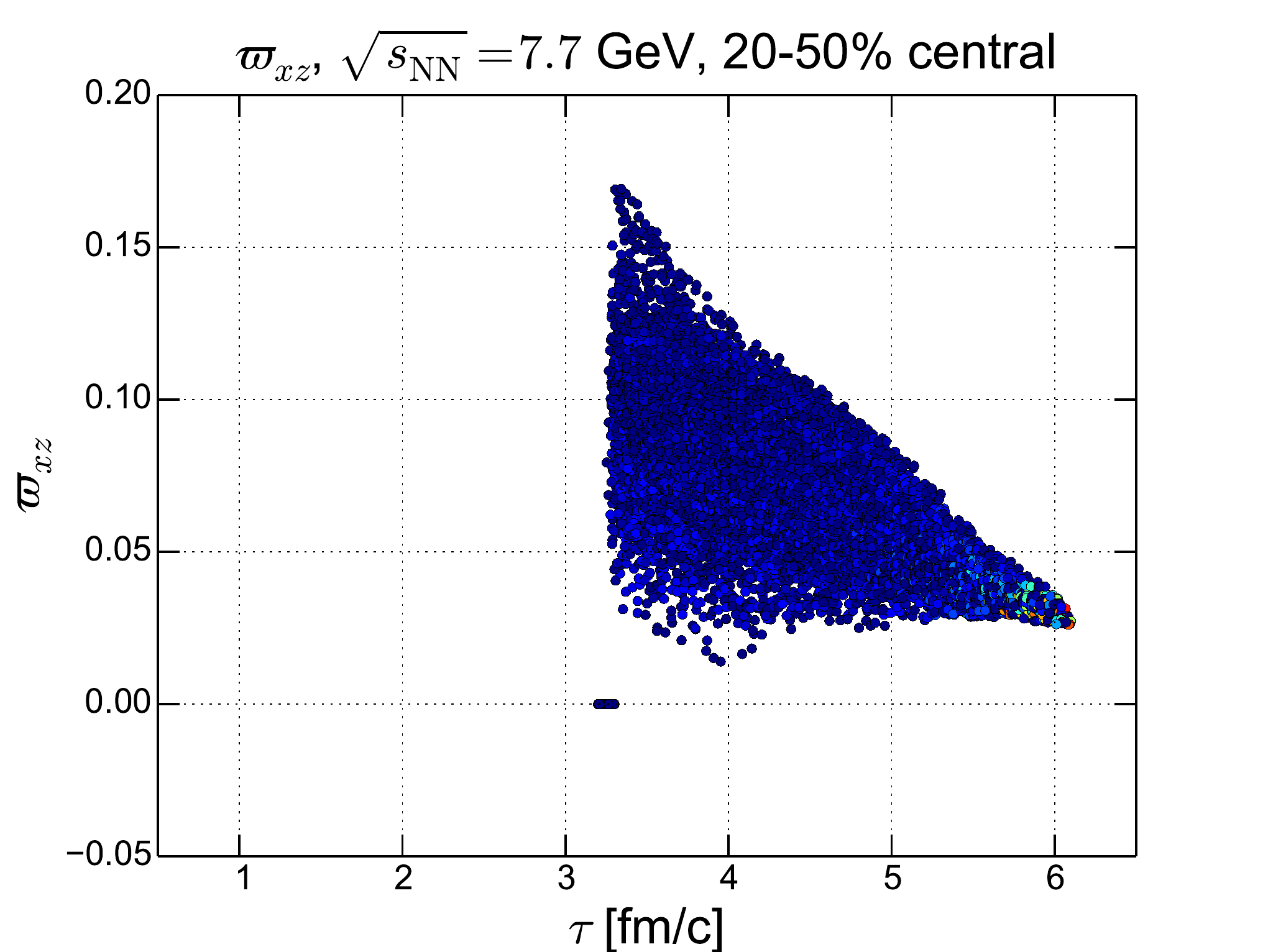}\\
\includegraphics[width=0.46\textwidth]{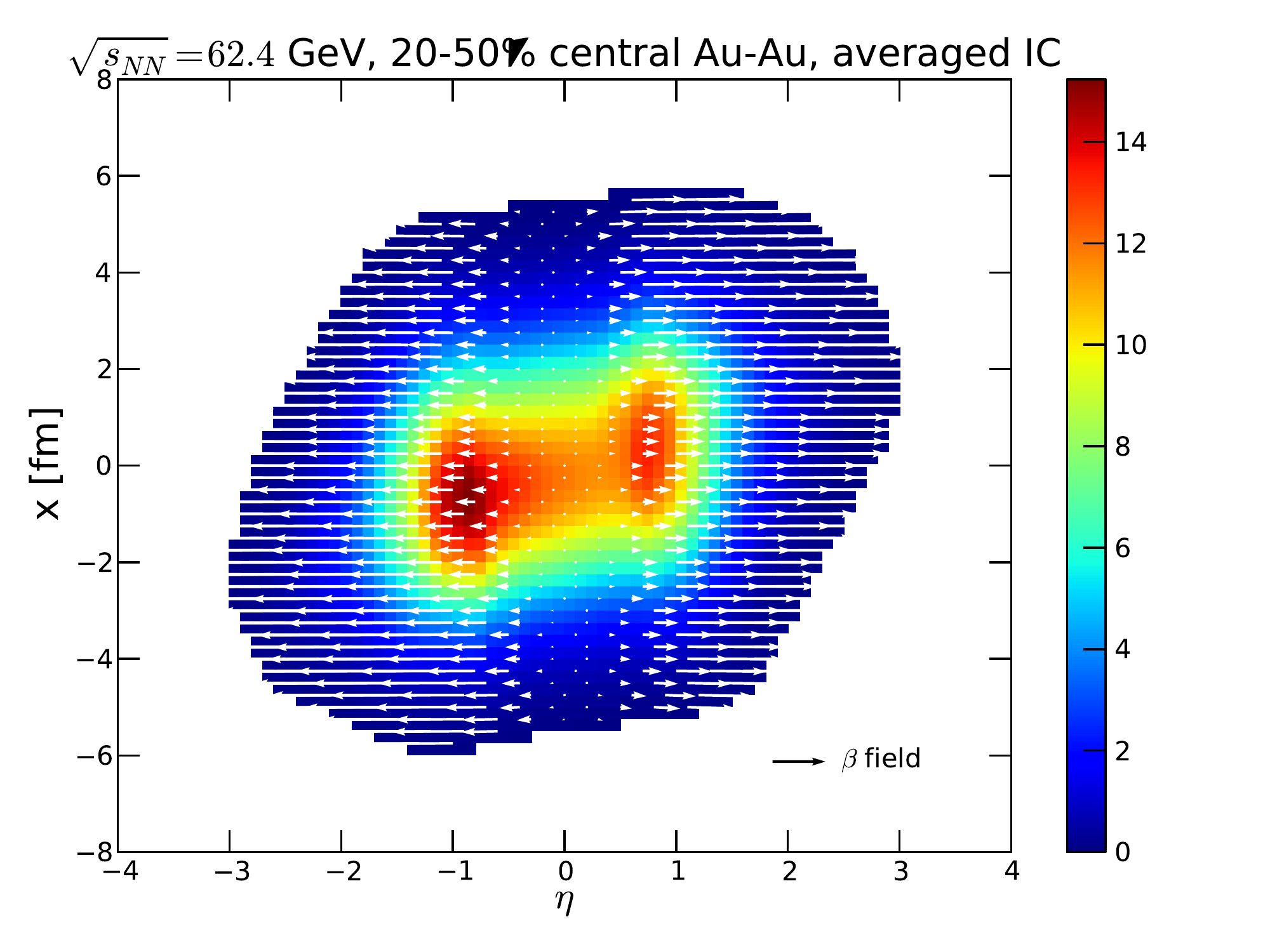}
\includegraphics[width=0.47\textwidth]{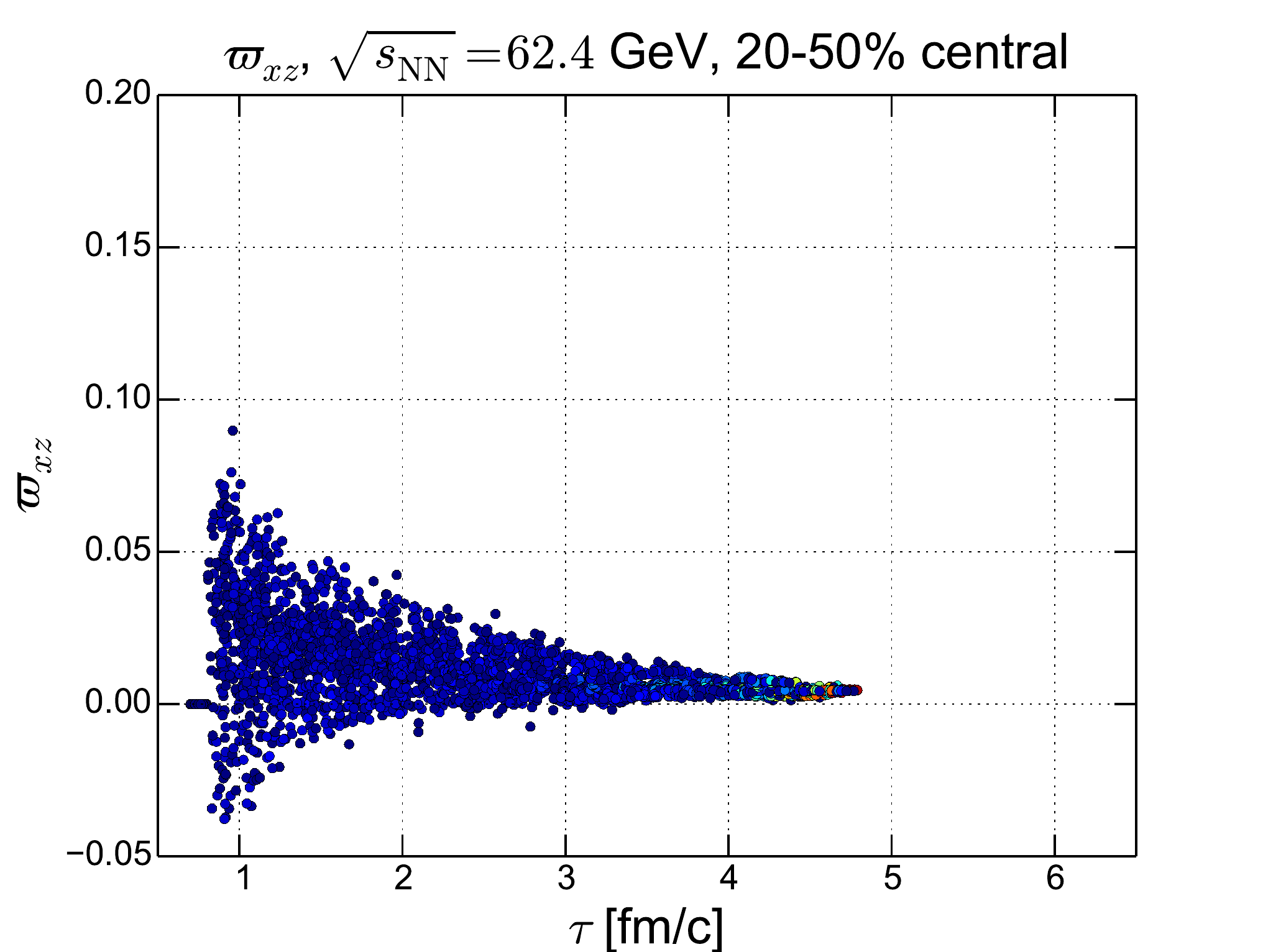}
\caption{Initial energy density profiles for hydrodynamic stage with arrows depicting 
initial four-temperature field superimposed (left column) and $\varpi_{xz}$ over space-time rapidity 
$|y|<0.3$ slice of particlization surface, projected onto time axis (right column). 
The hydrodynamic evolutions start from averaged initial state corresponding to 20-50\% 
central Au-Au collisions at $\sqrt{s_{\rm NN}}=7.7$ (top row) and $62.4$~GeV (bottom row).}
\label{fig-omegaXZ-tau}
\end{figure*}
%------------------------------------------------------------------------------------------

Now we have to understand the excitation function of the $p_T$ integrated $P_J$ which 
is calculated in the model. As it has been mentioned, $P_J$ at low momentum (which contributes 
most to the $p_T$ integrated polarization) has a dominant contribution proportional 
to $\varpi_{xz}p_0$. It turns out that the pattern and magnitude of $\varpi_{xz}$ over 
the particlization hypersurface change with collision energy.

We demonstrate this in Fig.~\ref{fig-omegaXZ-tau} for two selected collision energies. 
For this purpose we ran two single hydrodynamic calculations with averaged initial 
conditions from 100 initial UrQMD simulations each. At $\sqrt{\sNN}=62.4$~GeV, because 
of baryon transparency effect, the $x,z$ components of four-temperature vector around zero space-time rapidity are 
small and do not have a regular pattern, therefore the distribution of $\varpi_{xz}$ 
in the hydrodynamic cells close to particlization energy density includes both positive 
and negative parts, as it is seen on the corresponding plot in the right column. At
 $\sqrt{\sNN}=7.7$~GeV, baryon stopping results in a shear flow structure, which leads 
to same (positive) sign of the $\varpi_{xz}$.

In the right column of Fig.~\ref{fig-omegaXZ-tau}, we plot the corresponding $\varpi_{xz}$ 
distributions over the particlization hypersurfaces projected on the proper time axis. 
Generally speaking, hydrodynamic evolution tends to dilute the initial vorticities. One 
can see that longer hydrodynamic evolution at $\sqrt{\sNN}=62.4$~GeV in combination with 
smaller absolute value of average initial vorticity results in factor 4-5 smaller average 
absolute vorticity at late times for $\sqrt{\sNN}=62.4$~GeV than for $\sqrt{\sNN}=7.7$~GeV. 
This results in corresponding difference in the momentum integrated polarization at 
these two energies, that is mostly determined by low-$p_T$ $\Lambda$ which are preferentially 
produced from the Cooper-Frye hypersurface at late times.

%*********************************************************
\section{Feed-down and post-hadronization interactions}\label{feeddown}
%*********************************************************

%--------------------------------------------------------------------------------------------------------
\begin{figure}\label{fig-Pixy-resCorr}
\includegraphics[width=0.5\textwidth]{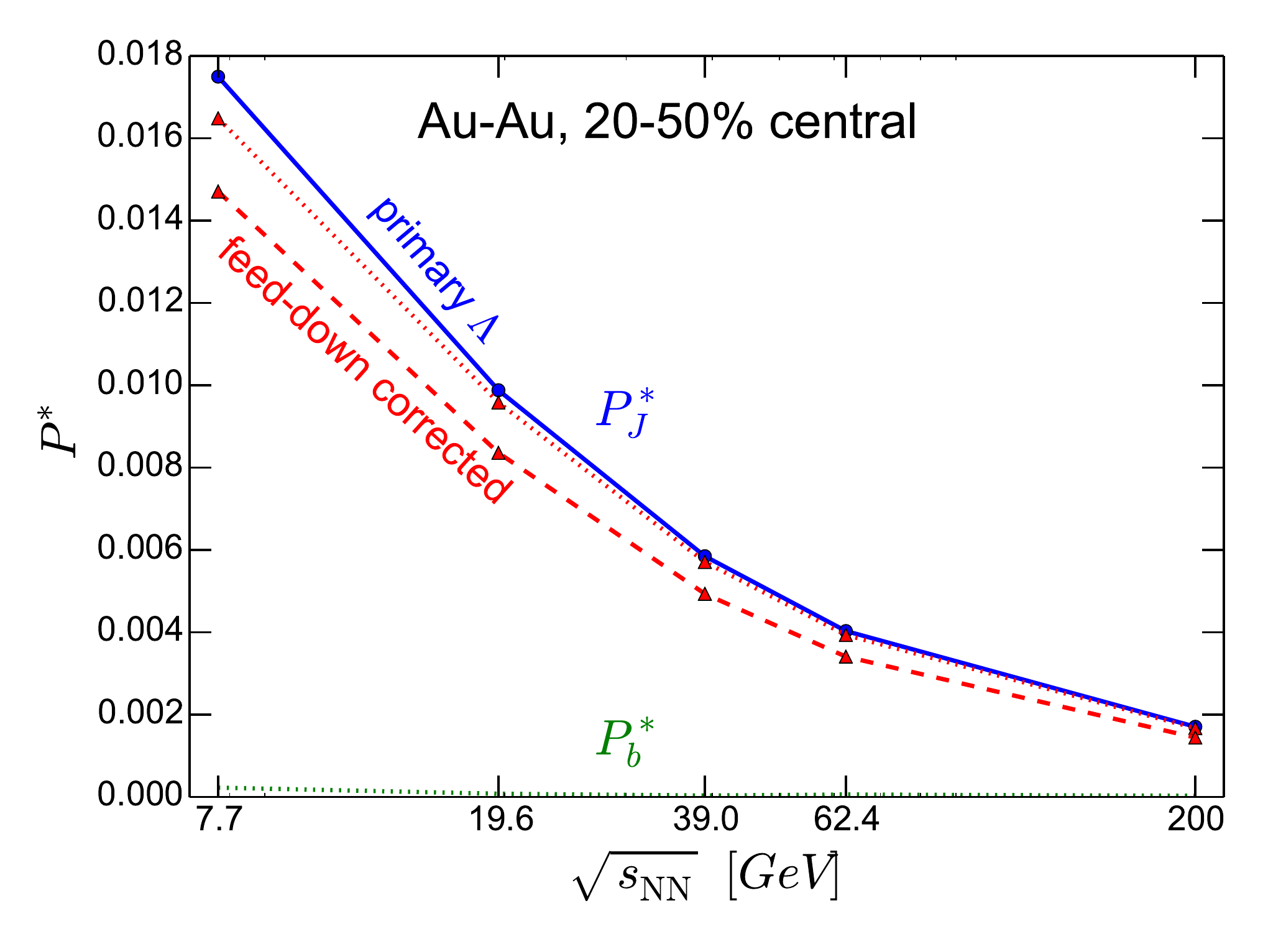}
\caption{Same as Fig.~\ref{fig-Pixy-sqrts}, dotted curve corresponds to polarization of feed-down corrected 
$\Lambda$ from $\Sigma(1385)$ and $\Sigma^0$ decays only. Dashed curve corresponds to feed-down corrected 
$\Lambda$ from $\Sigma^0$, $\Sigma(1385)$, $\Lambda(1405)$, $\Lambda(1520)$, $\Lambda(1600)$, $\Sigma(1660)$ 
and $\Sigma(1670)$, including cascade decays, e.g.\ $\Lambda(1405)\rightarrow\Sigma^0\rightarrow\Lambda$.}
\end{figure}
%--------------------------------------------------------------------------------------------------------

Thus far, we have calculated the polarization of the $\Lambda$'s produced from the 
plasma at the particlization stage - henceforth denoted as primary $\Lambda$ - when 
the fluid decouples at or right after hadronization. However, a sizable amount of 
the final $\Lambda$'s are products of resonance decays. Indeed, as long as one is 
interested in the {\em mean}, momentum-integrated, spin vector in the rest frame, 
it can be shown that a simple linear rule applies \cite{bkluv} that is:
\be\label{linear}
  {\bf S}^*_D = C {\bf S}^*_X
\ee
where $D$ is the daughter particle, X the parent and $C$ a coefficient whose expression 
may or may not depend on the dynamical decay amplitudes. If the coefficient $C$ does 
not depend on the dynamical decay amplitudes, it takes on rational values depending 
on Clebsch-Gordan coefficients, the initial values of spin and parity \cite{bkluv}. 
The values which are relevant for our calculation in various strong/electromagnetic decays 
with a $\Lambda$ or a $\Sigma$ hyperon in the final state are reported in table~\ref{tb-decayfactor}; 
for the full derivation of the $C$ coefficients see ref.~\cite{bkluv}.

A large fraction of secondary $\Lambda$'s comes from the strong $\Sigma(1385)\rightarrow\Lambda\pi$ and the 
electromagnetic $\Sigma^0 \rightarrow \Lambda \gamma$ decays \footnote{We denote $\Sigma(1385)$ below as 
$\Sigma^*$ for brevity.}. We found that - in our code - the fractions of primary $\Lambda$, $\Lambda$'s 
from $\Sigma^*$ decays and $\Lambda$'s from decays of primary $\Sigma^0$'s are respectively 28\%, 32\% and 17\%, with a 
negligible dependence on the collision energy. This is very close to the fractions extracted from a recent analysis \cite{becastein} 
within the statistical hadronization model: 25\%, 36\% and 17\%. The remaining 23\% of $\Lambda$'s consists of 
multiple smaller contributions from decays of heavier resonances, the largest of which are $\Lambda(1405)$, 
$\Lambda(1520)$, $\Lambda(1600)$, $\Sigma(1660)$ and $\Sigma(1670)$. Some of these resonances produce $\Lambda$'s 
in cascade decays, for example $\Lambda(1405)\rightarrow\Sigma^0\pi,\Sigma^0\rightarrow\Lambda\gamma$.

We start with the contribution from $\Sigma^*$, which is a $J^\pi=\sfrac{3}{2}^+$ state. In this case the factor 
$C$ in eq.~(\ref{linear}) is $1/3$ (see table~\ref{tb-decayfactor}) and, by using eq.~(\ref{pixpgen})
with $S=3/2$, we obtain that the mean spin vector of primary $\Sigma^*$ is 5 times the one of primary 
$\Lambda$. Thus, the mean spin vector of $\Lambda$ from $\Sigma^*$ decay is:
$$
  {\bf S}^{*} = \frac{1}{3} {\bf S}^*_{\Sigma^*} = \frac{5}{3} {\bf S}_{\Lambda,{\rm prim}}^{*}
$$
Similarly, for the $\Sigma^0$, which is a $\sfrac{1}{2}^+$ state, the coefficient $C$ is $-1/3$ 
(see table~\ref{tb-decayfactor}) and:
$$
  {\bf S}^{*} = - \frac{1}{3} {\bf S}_{\Sigma^0}^{*} = - \frac{1}{3} {\bf S}_{\Lambda,{\rm prim}}^{*}
$$
as primary $\Sigma^0$ are expected to have the same polarization as $\Lambda$.

More generally, the law (\ref{linear}) makes it possible to calculate the mean spin vector
inherited by the $\Lambda$ hyperons also in multi-step two-body decays decays with a simple 
linear propagation rule. We have thus computed the mean spin vector of primary + feed-down
$\Lambda$ as follows:
\begin{eqnarray}\label{eq-res-correction}
  {\bf S}^*= &&\frac{N_\Lambda + \sum\limits_{X}N_X\left[ C_{X\rightarrow\Lambda}b_{X\rightarrow\Lambda} -
 \frac{1}{3} C_{X\rightarrow\Sigma^0}b_{X\rightarrow\Sigma^0}\right]}
 {N_\Lambda + \sum\limits_{X}b_{X\rightarrow\Lambda}N_{X} + \sum\limits_Xb_{X\rightarrow\Sigma^0}N_X}
 \nonumber \\
 && \times {\bf S}^*_{\Lambda,{\rm prim}},
\end{eqnarray}
where the sum goes over resonances $X$ decaying into $\Lambda$ or $\Sigma^0$ and a pion, $N_X$ are 
the primary multiplicities of resonances, $C_{X\rightarrow\Lambda}$ and $C_{X\rightarrow\Sigma^0}$ 
are polarization transfer coefficients, $b_{X\rightarrow\Lambda}$ and $b_{X\rightarrow\Sigma^0}$ are 
the branching ratios for decay channels yielding in $\Lambda$ and $\Sigma^0$ respectively. In 
eq.~\ref{eq-res-correction} we have used the fact that nearly all $\Sigma^0$ decay to $\Lambda \gamma$ 
with polarization transfer $-1/3$, which allows to treat cascade decay contributions $X\rightarrow\Sigma^0\rightarrow\Lambda$. 
The results are shown in Fig.~\ref{fig-Pixy-resCorr},
where the thin dotted line corresponds to feed-down contributions $X=\Sigma^0,\Sigma(1385)$ only. 
Surprisingly, in this case the interplay of hadron chemistry and polarization transfer in the decays 
result in a correction factor, which varies between $0.94-0.98$ in the whole collision energy range. 
When we take all aforementioned resonances into account ($X=\Sigma^0$, $\Sigma(1385)$, $\Lambda(1405)$, 
$\Lambda(1520)$, $\Lambda(1600)$, $\Sigma(1660)$, $\Sigma(1670)$), using polarization transfer coefficients 
listed in table~\ref{tb-decayfactor}, we obtain the dashed line in Fig.~\ref{fig-Pixy-resCorr}, 
corresponding to a 15\% suppression of the mean polarization of $\Lambda$ with
respect to the primary polarization. This decrease is mostly due to the increase of the denominator 
in eq.~\ref{eq-res-correction} from the heavier resonance contributions, whereas their contributions 
to the numerator have opposite signs because of alternating signs of the polarization transfer coefficients.
%-----------------------------------------------------------------------------------------
\begin{table}
\vspace{10pt}
\begin{tabular}{|c|c|}
\hline
 Decay & $C$    \\ \hline
  $\sfrac{1}{2}^+ \to \sfrac{1}{2}^+ \;\; 0^- $    &   $-1/3$     \\ \hline
  $\sfrac{1}{2}^- \to \sfrac{1}{2}^+ \;\; 0^- $    &     1        \\ \hline
  $\sfrac{3}{2}^+ \to \sfrac{1}{2}^+ \;\; 0^- $    &   $1/3$      \\ \hline
  $\sfrac{3}{2}^- \to \sfrac{1}{2}^+ \;\; 0^- $    &   $-1/5$     \\ \hline
  $\Sigma^0 \to \Lambda \gamma $                   &   $-1/3$     \\ \hline
 \end{tabular}
\caption{Polarization transfer coefficients $C$ (see eq.~(\ref{linear})) to the $\Lambda$ or $\Sigma$ 
hyperon (the $\sfrac{1}{2}^+$ state) for various strong/electromagnetic decays.}
\label{tb-decayfactor}
\end{table}
%----------------------------------------------------------------------------------------------

There is, however, a further correction which is much harder to assess, i.e. post-hadronization interactions. In
fact, hadronic elastic interaction may involve a spin flip which, presumably, will randomize the spin direction 
of primary as well as secondary particles, thus decreasing the estimated {\em mean} global polarization in 
fig.~(\ref{fig-Pixy-resCorr}).
Indeed, in UrQMD cascade which is used to treat interactions after particlization, cross sections of $\Lambda$ 
and $\Sigma^0$ with most abundant mesons and baryons - calculated with the Additive Quark Model - are comparable 
to those of nucleons \cite{Bleicher:1999xi}. This implies that $\Lambda$'s do rescatter in the hadronic phase, 
and indeed we observe from the full cascade+hydro+cascade calculation that in the RHIC BES range only 10-15\% of 
primary $\Lambda$'s escape the system with no further interactions\footnote{The remaining 85-90\% of $\Lambda$ 
contain decay products of primary $\Sigma^0$, $\Sigma^*$ and other resonances up to $\Sigma(1670)$, which is covered by the calculations above.}, 
until they decay into pion and proton far away from the fireball. For the present, we are not able to provide
a quantitative evaluation of the rescattering effect on polarization, whose assessment is left to future studies.
The only safe statement for the time being is that the dashed line in fig.~(\ref{fig-Pixy-resCorr}) is an upper
bound for the predicted mean global $\Lambda$ polarization within the hydrodynamical model with the specific
initial conditions used in our calculation.

%**********************************************************************
\section{Conclusions}\label{sect-conclusions}
%**********************************************************************

In summary, we have calculated the global polarization of $\Lambda$ hyperons produced at midrapidity in Au-Au collisions 
at RHIC Beam Energy Scan collision energies, $\sqrt{\sNN}=7.7-200$~GeV, in the framework of 3 dimensional 
event-by-event viscous hydrodynamic model (\texttt{UrQMD+vHLLE}). The in-plane components of the polarization vector 
as a function of transverse momentum are found to have a quadrupole structure (similar to the one obtained in 
\cite{beca2015}) and can be as large as several percents for large transverse momentum. The mean, 
momentum integrated polarization vector is directed parallel to the angular momentum of the fireball and its 
magnitude substantially \textit{increases} from 0.2\% to 1.8\% as collision energy \textit{decreases} from full 
RHIC energy down to $\sqrt{\sNN}=7.7$~GeV. Such increase is related to (1) emerging shear flow pattern in beam 
direction at lower collision energies related to baryon stopping, and (2) shorter lifetime of fluid phase, which 
does not dilute initial vorticity as much as it does at higher collision energies. At the same time, we did not 
observe a linear relation between the polarization and the ratio angular momentum/energy of the fireball.

Significant fraction of the produced $\Lambda$ originate from resonance decays. We have calculated 
the contribution to $\Lambda$ polarization stemming from the decay of polarized heavier states 
and particularly for the two leading contributing channels: $\Sigma^0 \to \Lambda \gamma$ and $\Sigma(1385) 
\to \Lambda \pi$. It turns out that the contributions from $\Sigma^0$ and $\Sigma(1385)$ alone change 
the resulting polarization by few percents, whereas extending the list of feed-down corrections 
up to mass 1670~MeV hyperons results in about 15\% decrease of the resulting $\Lambda$ polarization.

It should be pointed out that we did not include post-hadronization rescatterings in our calculation, which are
likely to reduce the mean global polarization estimated in this work.

The calculated excitation function of the mean polarization reproduces the trend reported in the preliminary 
data from STAR \cite{MLisaTalk,upsal}, although the magnitude is about twice smaller, even without including
the further reduction due to hadronic rescattering. As the uncertainty on the parameters of the model cannot 
compensate for this discrepancy, this result suggests that a revision of the initial state model, particularly 
of the initial longitudinal flow velocity profile is necessary (see e.g. \cite{csernaivort}) as the present one 
provides insufficient magnitude of shear longitudinal flow to approach the experimentally observed magnitude of 
the polarization.

%******************************************************************
\section*{Acknowledgments}
%******************************************************************

We acknowledge useful discussions with M. Lisa and S. Voloshin.
This work was partly supported by the University of Florence grant 
{\em Fisica dei plasmi relativistici: teoria e applicazioni moderne}. 
The simulations have been performed at the Center for Scientific Computing 
(CSC) at the Goethe-University Frankfurt.

%====================================================================================================

%==============================================================================================

\end{document}